\documentclass[journal=jpccck,manuscript=article]{achemso}
\usepackage{graphicx}% Include figure files
\usepackage{dcolumn}% Align table columns on decimal point
\usepackage{bm}% bold math
\usepackage{textcomp}
\SectionNumbersOn
\setkeys{acs}{usetitle = true}

\usepackage{amsmath} 
\usepackage{amssymb} 
 \usepackage{amsfonts}

\usepackage{array}

\usepackage{multirow}

\usepackage{natbib}
\usepackage{mciteplus}

\author{Mila Miletic}
\affiliation{Research Group Simulations of Energy Materials, Helmholtz-Zentrum Berlin, Hahn-Meitner-Platz 1, D-14109 Berlin, Germany}
\alsoaffiliation{Institut f{\"u}r Physik, Humboldt-Universit{\"at} zu Berlin, Newtonstr. 15, 12489 Berlin, Germany}
 
\author{Karol Palczynski}
\affiliation{Research Group Simulations of Energy Materials, Helmholtz-Zentrum Berlin, Hahn-Meitner-Platz 1, D-14109 Berlin, Germany}

 \author{Matheus R. Jacobs}
\affiliation{Physics Department and IRIS  Adlershof, Humboldt-Universit\"at zu Berlin, Berlin, Germany}

\author{Ana M. Valencia}
\affiliation{Physics Department and IRIS  Adlershof, Humboldt-Universit\"at zu Berlin, Berlin, Germany}

\author{Caterina Cocchi}
 \email{Caterina.Cocchi@physik.hu-berlin.de}
\affiliation{Physics Department and IRIS  Adlershof, Humboldt-Universit\"at zu Berlin, Berlin, Germany}

\author{Joachim Dzubiella}
\email{joachim.dzubiella@helmholtz-berlin.de}
\affiliation{Physikalisches Institut, Albert-Ludwigs-Universit\"at Freiburg, Hermann-Herder Stra{\ss}e 3, D-79104 Freiburg, Germany}
\alsoaffiliation{Research Group Simulations of Energy Materials, Helmholtz-Zentrum Berlin, Hahn-Meitner-Platz 1, D-14109 Berlin, Germany}

\title{Impact of Polarity on the Anisotropic Diffusion of Conjugated Organic Molecules on the $\left(10\overline{1}0\right)$ Zinc Oxide Surface}
\keywords{Surface Transport, Anisotropic Diffusion, Hybrid Inorganic-Organic Semiconductors, Molecular Dynamics, Conjugated Organic Molecule, Hexaphenyl}

\begin{document}

\newpage
\begin{abstract}
We study the influence of polarity on the binding and diffusion of single conjugated organic molecules on the inorganic (10\=10) zinc oxide surface by means of all-atom molecular dynamics simulations at room temperature and above. 
In particular, we consider the effects of partial fluorination of the {\it para}-sexiphenyl ({\it p}-6P) molecule with chemical modifications of one head group ({\it p}-6PF2) or both (symmetric) head and tail  ({\it p}-6PF4).  Quantum-mechanical and classical simulations both result in consistent and highly distinct dipole moments and densities of the fluorinated molecules, which interestingly lead to a weaker adhesion to the surface than for {\it p}-6P.  The diffusion for all molecules is found to be normal and Arrhenius-like for long times. 
Strikingly, close to room temperature the polar molecules diffuse 1-2 orders of magnitudes slower compared to the {\it p}-6P reference in the apolar $x$-direction of the electrostatically heterogeneous surface, while in the polar $y$-direction they diffuse 1-2 orders of magnitude faster. We demonstrate that this rather unexpected behavior is governed by a subtle electrostatic anisotropic mismatch between the polar molecules and the chemically specific surface, as well as by increased entropic contributions coming from orientational and internal degrees of freedom.
\end{abstract}

\newpage
%\maketitle

%\tableofcontents
\maketitle
\section{\label{sec:intro}Introduction}
Hybrid structures of organic molecules and inorganic semiconductors (HIOS) combine the favorable properties of each material class into conjugates with enormous application potential.~\cite{1367-2630-10-6-065010, 0953-8984-20-18-184008, 0953-8984-22-8-084024}
The organic parts in particular offer a vast diversity in terms of chemistry and structure.~\cite{Marcon2008, Kirkpatrick2008, doi:10.1021/cr030070z, PhysRevLett.98.227402, C4CP04048A, Pithan2015, Moser2011, Krause2003, Zen2004} 
On top of that, the chemical composition and by association the structure are relatively easy to manipulate. 
By changing the structure, the opto-electronic properties of the organic materials can be fine-tuned in many subtle ways.
Combined with the influence of the inorganic parts, such as substrates onto which the organic molecules are deposited as thin films, HIOS are devices with well-tailored properties that cannot be realized with either material class alone.
In order to generate the desired structures in the thin films, it is critical to understand processes such as the molecular attachment to the inorganic surface and the diffusion of the molecules on the surface, which are relevant in the early stages of thin film growth.
For this to do, it is necessary to understand the structure of the respective molecules and the interactions between the inorganic and organic parts at the hybrid interface during interface formation. 

Early experiments were limited to the study of surface diffusion of single atoms \cite{Graham1975, Kellogg1994, Kellogg1985, Bassett1983}.
With recent advances in experimental techniques, many interesting features of internal molecular structure and surface kinetics have been revealed.
Adsorbates composed of more than one atom were demonstrated to have very different diffusion mechanisms due to an increased number of rotational and translational degrees of freedom \cite{Weckesser1999, Weckesser2001, doi:10.1021/jacs.5b08001}, compared to single atom adsorbates.    
With their capability to resolve microscopic details of diffusion that are still difficult to capture experimentally, molecular-scale simulations are necessary to interpret  experimental observations.

Important insights have been given in experimental and theoretical studies of adsorption and surface diffusion of the conjugated organic molecule {\it para}-sexiphenyl ({\it p}-6P) on different surfaces of the inorganic ZnO crystal.
Using X-ray diffraction measurements combined with atomic force microscopy, it was observed that the {\it p}-6P molecule adsorbs flat-lying on the surface with its long molecular axis (LMA) oriented perpendicular to the surface $[0001]$ direction (see Fig.~\ref{Figure_2}).~\cite{C004944C}
Studies based on the combination of first-principles and classical theoretical methods demonstrated that the energy of the ZnO/{\it p}-6P system is minimized when the molecule is oriented perpendicular to the $[0001]$ direction and that a high barrier exists for the translation of the molecule in the polar $[0001]$ direction.~\cite{PhysRevLett.107.146401}
This finding is explained by the influence of the electrostatic surface energy landscape which provides a template for the molecule to adsorb in a predefined fashion. 
Such a complex surface energy landscape was shown to impose anisotropic kinetic barriers for the adsorbed molecules.~\cite{doi:10.1021/jp507776h, doi:10.1021/jp203758p} 
The molecules then diffuse with high anisotropy in such systems. 
This, in turn, can lead to anisotropy in growth, e.g., one-dimensional clusters or clusters with a preferred attachment direction.~\cite{PhysRevE.98.042801, doi:10.1021/jp201343s, doi:10.1021/nn203105w} 
Hence, by smartly engineering the inorganic surfaces and the organic molecules, the diffusion and growth behavior can even be controlled towards device property optimization.
For example, recently it has been shown that after a subtle chemical modification of the the {\it p}-6P  molecule such as the replacement of two of the molecule's hydrogen atoms by fluorine atoms (i.e., fluorination), self-assembled structures on metal surfaces differ vastly from the original molecules.~\cite{doi:10.1021/acs.jpcc.8b03398}

Motivated by the substantial influence of fluorination observed in experiments, in this paper we study the diffusion of a single {\it p}-6P molecule and of two of its fluorinated derivatives, with chemical modifications of one head group ({\it p}-6PF2) or both (symmetric) head and tail ({\it p}-6PF4) on the inorganic (10\=10) zinc oxide surface. To provide a useful complement to experimental results, our aim is thus to investigate the effects of molecular polarity induced by fluorination on the single molecule structural properties, binding to the surface, as well as the anisotropic diffusion on the surface, that can provide new insights for future advanced self-assembly strategies.  Our study shows that for a hybrid system composed of the inorganic surface and single adsorbates that differ among each other based on the different number of polar groups, the prediction of the thermodynamics of adsorption and diffusion is not a trivial task and requires taking into account both the surface electrostatic landscape and polarity of the adsorbed molecules.  Furthermore, increasing the number of polar groups inside the organic molecule has different effects on the molecular surface diffusion in different (polar and apolar) directions of the surface. We show that the observed diffusion behavior is governed by an intricate balance of electrostatic attractive and repulsive interactions between the molecules and the underlying substrate and is substantially influenced by entropic contributions coming from the orientational and internal degrees of freedom of the adsorbed molecules.

%More text~\cite{Fox1950, Fox1955, Kanig1963, Mitsubishi, White2015}.
%Italic text Sun \textit{et al.}~\cite{Sun1994}, based on \textit{ab initio} calculations.
%Eposilon: $\varepsilon$ 

\section{\label{sec:methods}Methods} %%%%%%%%%%%%%%%%%%%%%%%%%%%%%%%%%%%%%%%%%%%%%%%%%%%%%%%%%%%%%%%%%%%%%%%%%%%%%%%%%%%%%%%%%%%%%%%%

\subsection{\label{sec:Force_field}Force fields and MD Simulations}
\begin{figure}[H]
\includegraphics[width=8.5cm]{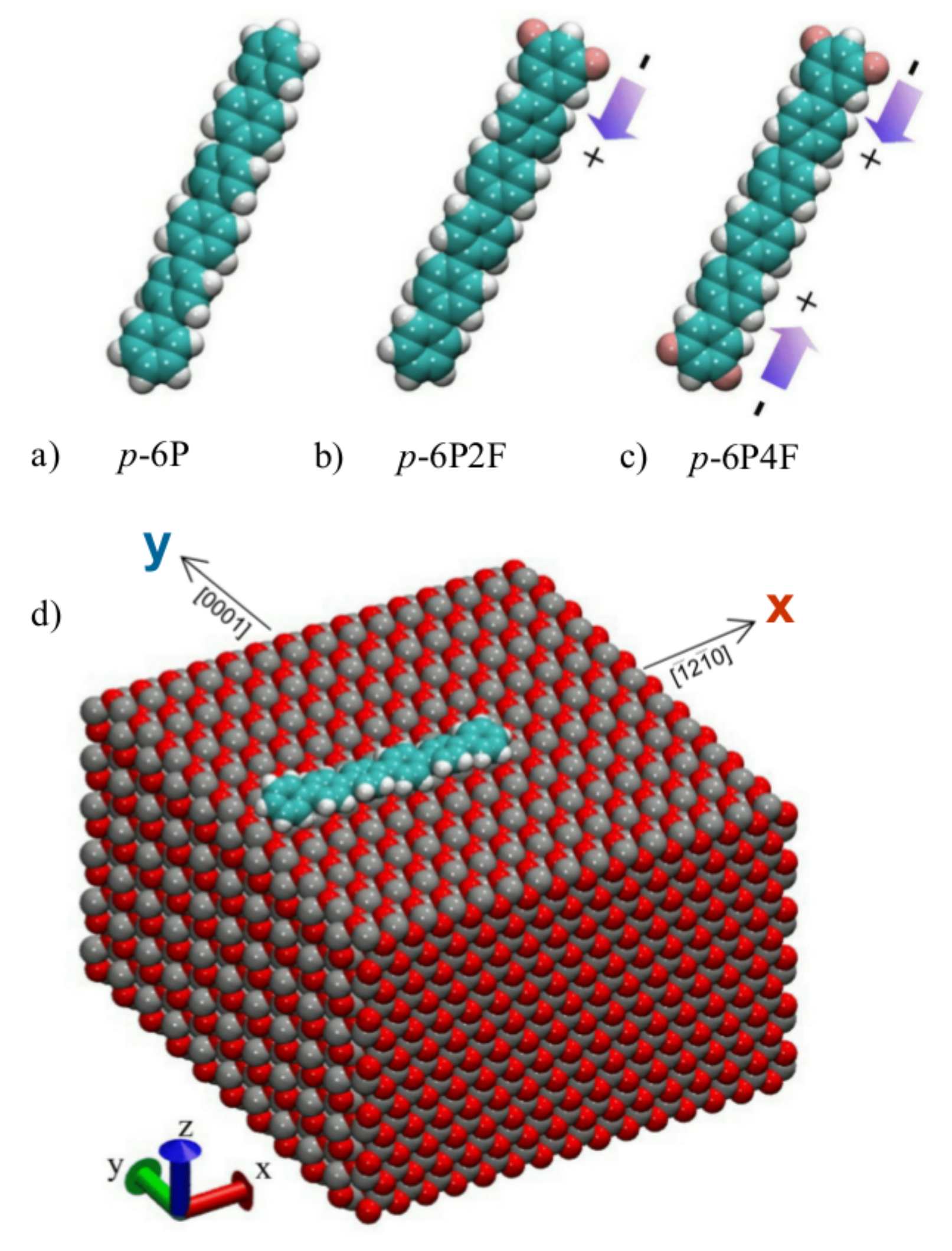}
\caption{a){\it Para}-sexiphenyl ({\it p}-6P) molecule and its b) asymmetrically {\it p}-6PF2 and c) symmetrically {\it p}-6PF4 fluorinated derivatives.
Cyan, white and pink colored spheres represent carbon, oxygen and fluorine atoms, respectively.
d) {\it p}-6P molecule adsorbed on the inorganic ZnO (10\=10) surface,
periodically repeated in the $x$- and $y$-directions.
Grey and red colored spheres represent zinc and oxygen atoms, respectively. Due to the underlying surface charge pattern, the molecule is, with a high probability, oriented perpendicular to the [0001] $y$-direction. \cite{PhysRevLett.107.146401, doi:10.1021/jp507776h}}
\label{Figure_2}
\end{figure}

In order to investigate the transport properties of the {\it p}-6P molecule and its symmetrically and asymmetrically fluorinated derivatives, {\it p}-6P2F and {\it p}-6P4F (Fig. \ref{Figure_2}), adsorbed on the ZnO (10\=10) surface, we perform classical, atomistic molecular dynamics (MD) simulations using the Gromacs simulation package (version 5.1).\cite{doi:10.1021/ct700301q} 
The interactions between all atoms are described by classical potentials. 
The Lennard-Jones (LJ) and Coulomb potentials are used for non-bonded interactions, rigid constraints (LINCS) are applied to the intramolecular bonds, and harmonic and cosine potential functions are used for the angular- and dihedral interactions, respectively. 
The force field parameters for the {\it p}-6P intramolecular potentials are taken from the general Amber force field (GAFF).\cite{JCC:JCC20035}  
The {\it p}-6P LJ and Coulomb potential parameters are taken from our previous simulation studies.\cite{doi:10.1021/cg500234r} 
It was demonstrated that with this force-field (Hamiltonian) the molecules self-assemble into the correct experimental room-temperature {\it p}-6P bulk crystal structure by simple temperature annealing, without any external bias.\cite{doi:10.1021/cg500234r} 
The LJ size and energy parameters for ZnO are taken from GAFF \cite{JCC:JCC20035} and the partial charges placed on the individual Zn and O atoms are taken from previous work. \cite{doi:10.1063/1.463056}
Our model system is comprised of a ZnO slab containing  $N_{x}$ $\times$ $N_{y}$ $\times$ $N_{z}$ = 15 $\times$ 10 $\times$ 6 unit-cells, periodically repeated in $x$- and $y$-directions with unit-cell parameters $L_{x} =0.329$~nm and $L_{y} =0.524$~nm taken from  previous work,\cite{doi:10.1021/jp507776h} and a single organic molecule adsorbed on the top of the ZnO (10\=10) surface.
The motion of an atom $i$ at position $\vec{r_{i}}$ is described by the Langevin equation of motion
\begin{equation}
m_i\frac{d^{2}\vec{r_{i}}}{dt^{2}} = -m_i\xi_i\frac{d\vec{r_{i}}}{dt}+\vec{F_{i}}+\vec{R_{i}}  ,  
\label{eq1}\end{equation}
where $m_i$ is the atomic mass, $\xi_i$ is the friction constant in units of ps$^{-1}$, $\vec{F_{i}}$ is the force acting on atom $i$ due to all other atoms, and $\vec{R_{i}}(t)$  is the random force obeying the usual fluctuation-dissipation theorem. \cite{doi:10.1021/ct700301q}
The equations of motion are integrated using a leapfrog algorithm with a time step of 2 fs. 
The surface atoms are fixed to their initial positions during the simulation, i.e., eq.~(\ref{eq1}) is not solved for the surface atoms resulting in a static surface potential.
In experimental reality, the molecule which is on the surface but otherwise surrounded by vacuum gets its heat from the vibrating surface through random collisions only, if we disregard any form of radiation flux.
Simultaneously, a part of the heat dissipates back to the surface via friction.  
The equilibrium between incoming and outgoing flux then determines the actual temperature of the molecule.
In the canonical-ensemble (NVT) simulations, we have to make sure that the heat balance of the molecule is faithfully reproduced, even if all the surface atoms are frozen (that is, static).
Thus, we couple the molecule in eq.~(\ref{eq1}) to an artificial heat bath with the bath friction $\xi=1$~ps$^{-1}$ as an auxiliary method to make sure the molecule remains at a constant temperature. 
The long range Coulomb interactions in the system are computed by the Particle Mesh Ewald (PME) method,~\cite{doi:10.1021/ct700301q} using a Coulomb cutoff distance of 1 nm with interpolation order 4 and 30 $\times$ 20 $\times$ 35 grid points in $x$, $y$ and $z$ directions. 
For the vdW interactions a cutoff of 1.0 nm was applied.

\subsection{Analysis of surface binding and diffusion}

\subsubsection{\label{sec:Binding}Single molecule surface binding}
We calculate the molecule/ZnO potential energy of binding, the binding free energy and the entropy of binding for the three investigated molecules.  
We define the potential energy of binding as the difference between the total energies of the system when the molecule is adsorbed on the substrate compared to when it is far away from the substrate:
\begin{equation}
\Delta U_{b} = U_{bind}^{ZnO+mol} = U_{tot}^{ZnO+mol} - (U_{tot}^{ZnO} + U_{tot}^{mol}),
\label{U_bind}
\end{equation}
where $U_{bind}^{ZnO+mol}$ represents the binding energy to the surface, $U_{tot}^{ZnO+mol}$ the total energy of the molecule adsorbed on the surface, $U_{tot}^{ZnO}$ is the energy of the surface, and $U_{tot}^{mol}$ the energy of the single molecule.

The free energy for the binding/unbinding process, $\Delta F_{b}$,  was estimated from the potentials of mean force (PMF), as the difference between the highest and lowest values of the PMF curve. 
The PMF were calculated from steered MD simulations at $T = 525$~K, where the center-of-mass (COM) of the molecule is connected to a virtual site via a harmonic potential with the spring constant $k = 5000$~kJ~mol$^{-1}$~nm$^{-1}$, while the virtual site moves away from the surface with a constant velocity of $10^{-4}$~nm~ps$^{-1}$.
The PMF is then obtained from the integral of the net force (the bath friction force being subtracted) acting on the harmonic spring while the molecule is pulled away from the surface.  The entropic contribution to the free energy of binding can be calculated from $T\Delta S_{b} = \Delta U_{b} - \Delta F_{b}$, where $\Delta U_{b}$ is the binding energy to the surface from eq.~(\ref{U_bind}). 
The origin $z=0$ is defined as the $z$-coordinate of the COM of the Zn atom in the top-most layer of the surface. 

\subsubsection{\label{sec:Diffusion}Surface diffusion analysis}
The total one-dimensional long time diffusion coefficients in the $x$- and $y$-directions are obtained from the $x$- ($y$-) components of the mean squared displacements (MSD) of the molecular COM, respectively, via
\begin{equation}
\langle (x(t)-x(t_{0}))^{\rm 2} \rangle = \lim_{t \to \infty} 2 D_{x}^{\rm tot} t
\label{MSD}
\end{equation}
($y$ analogously). The total friction, that is imposed on the molecule during its diffusion on the surface, can be separated into two independent contributions. 
The first contribution, which is the one of interest, comes from the existence of surface atoms that interact with atoms of the molecule by the electrostatic and LJ interactions.  This is the molecule-surface friction. The second, artificial, contribution comes from the random force that we employ to maintain a full energy dissipation and partitioning among the system components. 
These two independent contributions are additive, i.e., $\xi_\alpha^{\rm tot}=\xi_\alpha^{mb}+\xi_\alpha^{ms}$, where $\alpha$=$x$, $y$ and $\xi_\alpha^{mb}$ and $\xi_\alpha^{ms}$ stand for molecule-bath and molecule-surface friction, respectively. 
The total friction coefficient $\xi_\alpha^{\rm tot}$ can be calculated from the usual Stokes-Einstein relation for Brownian surface diffusion:
\begin{equation}
 \xi_\alpha^{\rm tot}=\frac{k_{B}T}{MD_\alpha^{\rm tot}\left(T\right)},   
\label{friction}
\end{equation}
where $D_\alpha^{\rm tot}\left(T\right)$ are the $T$-dependent diffusion coefficients with $\alpha$=$x$, $y$, $M$ is the mass of the molecule and $k_{B}T$ is the thermal energy. 
With the known Langevin (bath) friction of 1 ps$^{-1}$, that is set as an input parameter for the simulation, $\xi_\alpha^{ms}$ simply follows as $\xi_\alpha^{ms}=\xi_\alpha^{\rm tot}-\xi_\alpha^{mb}$.
The reciprocal diffusion coefficient in eq.~(\ref{friction}) can also be divided in two parts as ${1}/{D_\alpha^{\rm tot}\left(T\right)}={1}/{D_\alpha^{mb}\left(T\right)}+{1}/{D_\alpha^{ms}\left(T\right)}$, where the desired molecule-surface diffusion in direction $\alpha$ follows as 
\begin{eqnarray}
\frac{1}{D_\alpha \left(T\right)}:=\frac{1}{D_\alpha^{ms}\left(T\right)}=\frac{1}{D_\alpha^{\rm tot}\left(T\right)}-\frac{1}{D_\alpha^{mb}\left(T\right)}.
\end{eqnarray}

\subsection{\label{sec:Free_en}Free energy landscapes}
In order to calculate the free energy landscapes of the molecules for diffusion on the surface, as well as to quantify the respective entropies, we extract from the simulation trajectories the (one-dimensional) equilibrium probabilities of spatial presence $P(x)$ and $P(y)$ for the molecular COM.  The free energies are obtained from the Boltzmann inversion 
\begin{equation}
F(x) = -k_{B}T \ln P(x)
\label{eq3}
\end{equation}
($y$ analogously). The free energy barriers, $\Delta F_{\alpha}$,  where $\alpha=x,y$, are then defined as the difference between the maximum and the minimum of $F(\alpha)$.
Since free energies are calculated at different temperatures $T$, the entropies can be obtained from $\Delta S_{\alpha}=-\partial\Delta F_{\alpha}(T)/\partial T$.
We calculate the derivative numerically using the finite-differences scheme generally written as $\Delta S_\alpha(T)\simeq-[\Delta F_\alpha(T+\Delta T)-\Delta F_\alpha(T-\Delta T)]/2\Delta T$ with $\Delta T =25$~K in our case.
The energy then follows from $\Delta E_{\alpha}=\Delta F_{\alpha}+T\Delta S_{\alpha}$.

\subsection{\label{sec:Dipoles}Molecular dipole moments}
We compute the dipole moments of the considered molecules in both classical MD simulations and in the quantum-mechanical framework provided by density-functional theory (DFT).~\cite{PhysRev.136.B864}  We calculate the total molecular dipole moment for the three investigated molecules for a given fixed molecular configuration with respect to the molecular COM, using the formula
\begin{equation}
\vec{\mu} = \sum\limits_{i=1}^{n} q_{i} (\vec{r}_{i}-\vec{d})
%\vec{\mu} = \sum\limits_{i=1}^{n} q_{i} (\vec{r_{i}}-\vec{d})
%\vec{\mu} = q \vec{d}
\label{eq4}
\end{equation}
where $n$ is the total number of atoms in the molecule, $\vec{r}_{i}$ is the position vector of the atomic partial charge $q_{i}$ and $\vec{d}$ is the position vector of the COM of the molecule. In classical MD the $q_i$ are simply given by the partial charges as defined in the input force field. We consider either a fully planar or the twisted minimum energy configuration of the molecules. Planar configurations represent the molecules adsorbed on a ZnO surface while twisted geometries correspond to the molecules in vacuum. 

In the DFT framework, electron energies and wave-functions are obtained from the solution of the Kohn-Sham (KS) equations~\cite{PhysRev.140.A1133}
\begin{equation}
\left[-\dfrac{1}{2}\nabla^2_{i} + v_{\rm eff}(\vec{r}) \right]\phi_{i}(\vec{r})=\epsilon_{i}\phi_{i}(\vec{r}),
\label{eq:ks}
\end{equation}
where $\phi_{i}(\vec{r})$ and $\epsilon_i$ are the KS eigenvalues and eigenenergies. 
On the left-hand-side of eq.~\eqref{eq:ks} we find the kinetic energy terms and the effective Kohn-Sham potential $v_{\rm eff}$ which is given by the sum of three terms:
\begin{equation}
v_{\rm eff}(\vec{r}) = v_{ext}(\vec{r}) + \int\frac{\rho(\vec{r'})}{|\vec{r}-\vec{r'}|}d\vec{r'} + v_{xc}(\vec{r}).
\label{eq:veff} 
\end{equation} 
The first term in eq.~\eqref{eq:veff} accounts for the electron-nuclear attraction, the second one is the Hartree potential, and the third one the exchange-correlation potential.
Here, we evaluate the latter term for this term using the generalized gradient approximation in the Perdew-Burke-Ernzerhof parameterization~\cite{PhysRevLett.77.3865}.
Carbon and fluorine 1$s$ states are treated as core electrons using the Hamann-Schlüter-Chiang-Vanderbilt pseudopotentials~\cite{PhysRevB.32.8412}. 
Calculations are performed with the \textsc{Octopus} code~\cite{doi:10.1002/pssb.200642067} adopting a grid with spacing 0.013~nm on a set of spheres of radius 0.4~nm centered around each atom.
Within the Born-Oppenheimer approximation, the nuclear degrees of freedom are optimized by minimizing the forces acting on them until they are smaller than 0.5~eV~nm$^{-1}$ ($\approx 48.24$~kJ~mol$^{-1}$~nm$^{-1}$).
To do so, we employ the so-called FIRE algorithm introduced and discussed in Ref.~\cite{PhysRevLett.97.170201}.
This step is necessary to ensure that the considered atomic configuration corresponds to a (local) minimum for DFT.

\section{\label{sec:results}Results and Discussion}

\subsection{Single molecule properties \label{single_mol}}
First, we compare the inter-ring torsional angles and energies, the length of the LMAs and the total dipole moments of the three molecules (see Table \ref{Table2}).
For this, we simulate each molecule separately in vacuum without the surface.
In each simulation, we constrain all five torsional angles in the molecules to the same absolute value, though with alternating sign, using dihedral restraints and sample the atomic coordinates and the total energy for a wide range of torsional angles.
%The total energy is comprised of intramolecular Coulomb and Lennard-Jones interactions in addition to the angular and dihedral potentials. 

The MD result shows a roughly parabolic $\Delta E(\phi_{\rm C-C})$ profile (see Fig. S1 of the Supporting Information) with the minimum energy at an angle of 29.5\textdegree~(Table \ref{Table2}). 
The energy difference between the planar and twisted states, $\Delta E_{\rm p-t}$, is compared in Table \ref{Table2} and together with the minimum energy angle values is well within the spread of previously published MD and quantum mechanical calculation results for the single {\it p}-6P molecule.~\cite{doi:10.1021/cg500234r}
\begin{table}
\caption{Comparison of structural and energetic properties of isolated {\it p}-6P, {\it p}-6P2F and {\it p}-6P4F molecules between planar structure and a twisted conformation with a minimum energy angle from the classical MD simulations. 
The length $L$ is the distance between the terminal carbon atoms, $\Delta$ $E_{p-t}$ is the energy difference between a planar and a twisted conformation, and $\phi_{\rm C-C}$ is the torsional angle at which the internal energy is minimal.}
\begin{tabular}{ccccc}
molecule & $L_{\rm planar}$ (nm) & $L_{\rm twisted}$ (nm) & $\Delta$ $E_{\rm p-t}$(kJ/mol) & $\phi_{\rm C-C}$\tabularnewline
\hline 
\hline 
p-6P & 2.47$\pm$0.00 & 2.45$\pm$0.01 & 25.54 & 30.2\tabularnewline

p-6P2F & 2.47$\pm$0.00 & 2.46$\pm$0.01 & 25.51 & 29.5\tabularnewline
 
p-6P4F & 2.47$\pm$0.00 & 2.45$\pm$0.02 & 25.17 & 29.7\tabularnewline
\hline 
\hline 
\label{Table2}
\end{tabular}
\end{table}
Next, in Table \ref{Table2} we compare the length of the long axis of the three molecules, in the fully planar configuration and in the twisted configuration.
The resulting values are very similar for all molecules and in satisfactory agreement with quantum mechanical calculation results for the single {\it p}-6P.~\cite{doi:10.1021/cg500234r} 
In the case of the molecules investigated here, the molecular twisted configuration was found to be energetically more favorable compared to the planar configuration in the gas-phase.
We conclude that the different polarity has no noticeable influence on the geometrical properties of the isolated molecules. 

Furthermore, we calculate the total molecular dipole moments for the three investigated molecules in respect to the molecular COM, since the partial fluorination introduces a local dipole moment along the LMA compared to the otherwise non-dipolar {\it p}-6P molecule. 
Both, MD and DFT calculations yield that the $x$-component of the dipole moment is indeed the most dominant one (where the $x$-axis is the axis parallel to the LMA) with vanishing dipole moment contributions in the perpendicular $z$ and $y$ directions. 
In Table~\ref{table:dipoles-DFT} we report the averaged $x$ components of the dipole moments, $\langle \mu_x \rangle$, computed by DFT and MD methods. While $\it p$-6P has no global dipole nor strong local ones, the fluorination induces a strong dipole moment at the end of the molecule. For the asymmetric $\it p$-6P2F this leads to a large global dipole, cf. Table~\ref{table:dipoles-DFT}, while in the anti-symmetric $\it p$-6P4F the two large end-dipoles cancel each other out and the total dipole vanishes. 

\begin{figure}[t!]
\includegraphics[width=10cm]{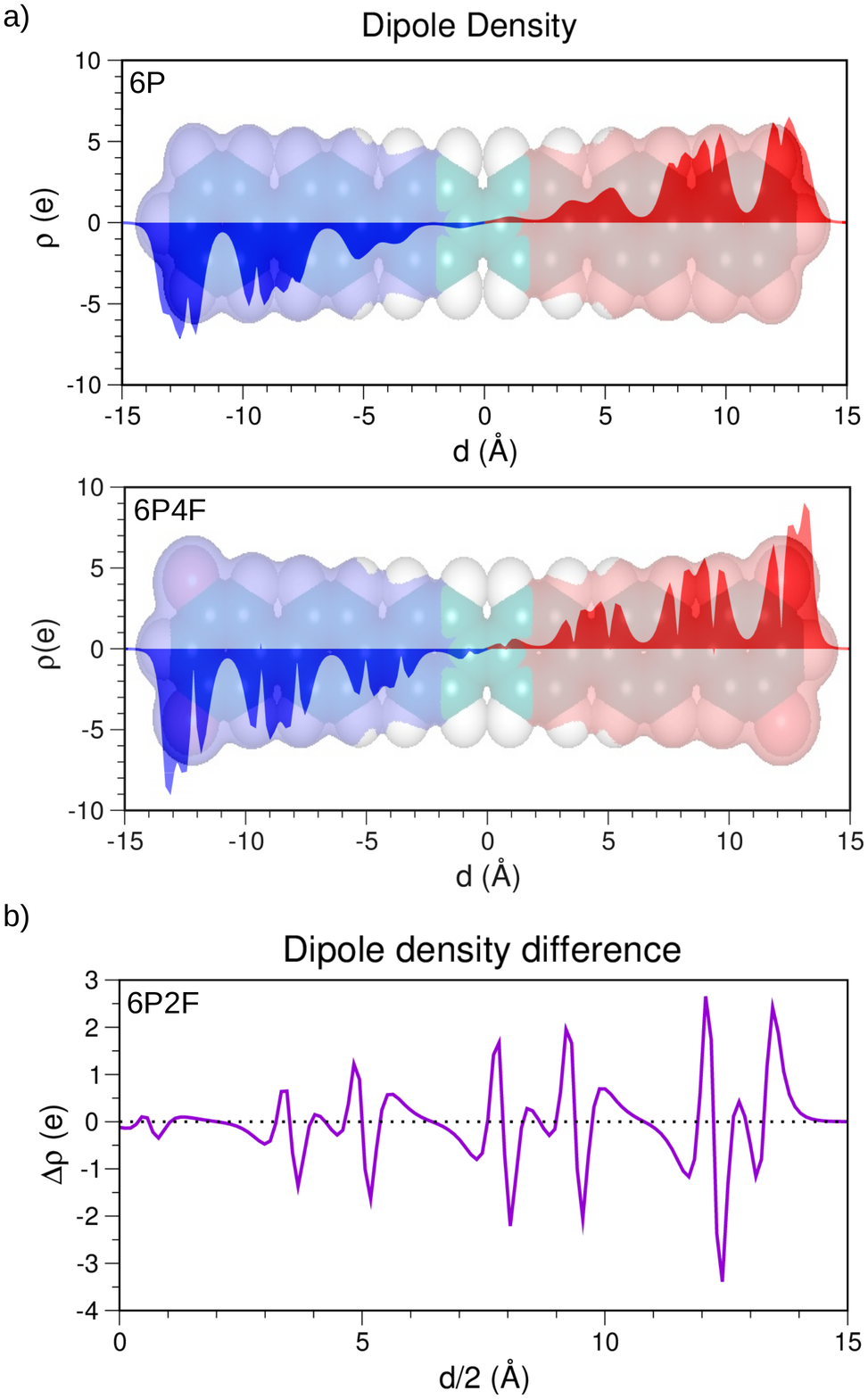}
\caption{a) Total dipole density along the long molecular axis ($x$-direction) of the symmetric twisted molecules {\it p}-6P and {\it p}-6P4F; b) Dipole density difference computed at each point of  one half of the {\it p}-6P2F molecule with respect to the mirror points in the other half along the $x$-axis: $d/2 = 0$ indicates the center of the molecule and $d/2 = 15$~\AA~its edge.}
\label{Dipole-dens}
\end{figure}
\begin{table}[ht]
\caption{Comparison of the $x$-component (in direction of the LMA) of the total molecular dipole moment (in Debye), computed by DFT and MD between the planar and twisted molecular configurations.}
\centering
\begin{tabular}{ c | c c c | c c c }
     &         &     planar  &     &      &   twisted   &\\
Molecule & {\it p}-6P & {\it p}-6P2F & {\it p}-6P4F & {\it p}-6P & {\it p}-6P2F & {\it p}-6P4F\\
\hline 
\hline
$\langle \mu_x^{MD} \rangle$  [D] & 0.00 & -2.43 & 0.03 & -0.02 & -2.69 & 0.01\\
$\langle \mu_x^{DFT} \rangle$ [D] & -0.01 & -2.88 & 0.04 & -0.01 & -2.97 & 0.01\\
\hline
\hline
\end{tabular}
\label{table:dipoles-DFT}
\end{table}
For details, in Fig.~\ref{Dipole-dens} the dipole density in units of electronic charge $e$ is shown.  In Fig.~\ref{Dipole-dens}(a), the results for {\it p}-6P and {\it p}-6P4F in their twisted geometries are reported.  Negative values of the dipole density are found in the left side of all molecules and positive values on the right side, with maximized values at the edges. There is a substantial increase of dipole density in {\it p}-6P4F at the ends due to fluorination. The relative change of local dipole density can be more clearly appreciated from Fig.~\ref{Dipole-dens}(b). There, we plot the charge difference across the two halves of the molecule {\it p}-6P2F: the charge at each point on  the right half is subtracted from the charge on the symmetric point on the left side. In the symmetric molecules {\it p}-6P and {\it p}-6P4F, such charge difference vanishes across the whole backbone. Conversely in the case of {\it p}-6P2F (Fig.~\ref{Dipole-dens}(b)) the asymmetry of the molecule gives rise to pronounced maxima and  minima increasing in magnitude toward the end of the oligomer. An analogous picture is obtained for the molecules in their planar geometry, as indicated in Table~\ref{table:dipoles-DFT}. In the {\it p}-6P2F molecule, a dipole moment of almost 3 Debye appears along $x$ in both the planar and twisted geometry, which exhibits remarkably similar values in the $x$-direction, while in $y$ and $z$ the dipole moments vanish. Similar behavior was found in graphene nanoflakes investigated from Hartree-Fock based methods, where the interplay between end functionalization and structural distortions was studied in detail.~\cite{doi:10.1021/jp300657k}

\subsection{Binding to the surface}
We continue by simulating each molecule on the inorganic ZnO [10\=10] surface.
The molecules are observed to adsorb in equilibrium in a flat-lying geometry on the surface, with their long molecular axis mostly aligned with the alternating rows of surface oxygen atoms along the $x$-axis. 
We determine the most probable torsional angles of 20.25\textdegree~for the {\it p}-6P, 21.56\textdegree~for the {\it p}-6P2F and 22.79\textdegree~for the {\it p}-6P4F, adsorbed on the surface.
The molecular head- and tail phenyl groups are recorded to have an increased rotational freedom compared to the inner phenyl groups with the most probable torsional angles of 22.92\textdegree, 24\textdegree~and 26\textdegree~for the head-groups of {\it p}-6P, {\it p}-6P2F and {\it p}-6P4F, respectively. 
If we compare these values with the values listed in Table \ref{Table2} for the isolated molecules, we can see that the molecular torsional freedom is reduced due to the interaction with the underlying substrate. 
We can also see that the head-groups of {\it p}-6P2F and {\it p}-6P4F display higher rotational freedom when on the substrate, compared to the head-group of the {\it p}-6P, indicating less strong adhesion.

\begin{figure}[ht]
\includegraphics[width=8.0cm]{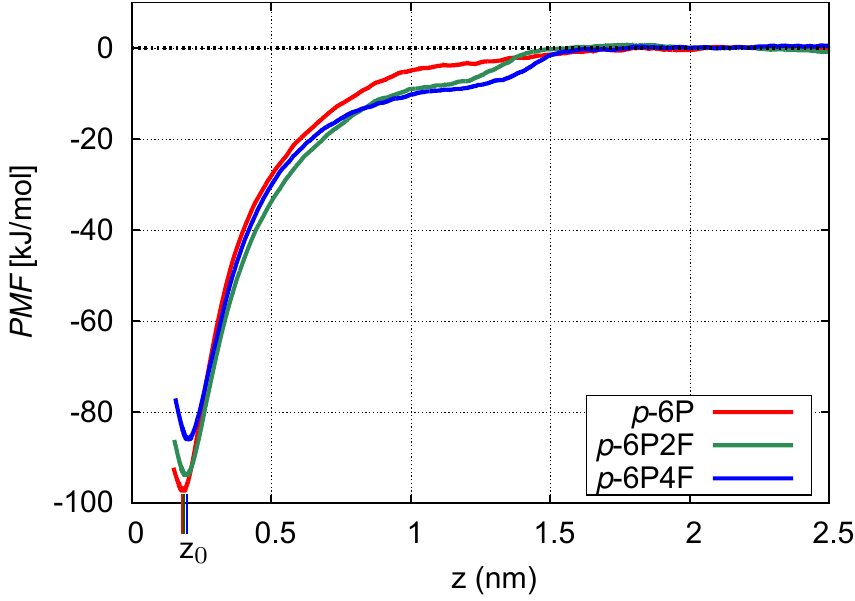}
\caption{PMF as function of distance from the surface in $z$-direction at $T$ = 525 K for {\it p}-6P, {\it p}-6P2F and {\it p}-6P4F. 
The minimum of the curves corresponds to the distance of 0.184 nm, 0.195 nm and 0.202 nm ($z_{0}$) for {\it p}-6P, {\it p}-6P2F and {\it p}-6P4F, respectively.
The origin $z$=0 is defined as the $z$-coordinate of the COM of the Zn atom in the top-most layer of the surface. The value of $\Delta F_{b} (z_{0})$ for the  binding/unbinding process is defined as the difference between the highest and lowest values of the PMF curve.}
\label{Fig_bind}
\end{figure}

As the next step, we calculate the molecule/ZnO surface binding free energy, entropy of binding and potential energy of binding for the three investigated molecules (see Methods \ref{sec:Binding}). 
The PMF for $T = 525$~K are plotted in Fig. \ref{Fig_bind}. 
The origin $z=0$ in Fig. \ref{Fig_bind} is defined as the $z$-coordinate of the COM of the Zn atom in the top-most layer of the surface. 
The minima of the curves, $z_{0}$, correspond to the distances of 0.184 nm, 0.195 nm and 0.202 nm for the {\it p}-6P, {\it p}-6P2F and {\it p}-6P4F, respectively.
The binding free energy is defined as the difference between the highest and lowest values of the PMF curve.

The results are presented in Table~\ref {Table_bind} alongside the corresponding energies and entropies of binding.
We find that the binding free energies and potential energies of binding decrease with an increase in molecular polarity, i.e., the more polar molecules bind less strong. Strikingly, the entropic penalty due the restrictions of configurational freedom on the surface is substantial and constitutes more than 1/3 of the total binding free energy. 
We find that the {\it p}-6P has a higher change in the entropy associated with binding than the other two molecules.
This again demonstrates that the overall attractive interactions between the molecule and the surface are stronger in the binding of the {\it p}-6P to the ZnO than for the other two molecules which have more configurational freedom on the surface. Apparently, repulsive electrostatic interactions between fluorine and ZnO cause the potential energy of binding to decrease compared to the non-fluorinated {\it p}-6P, i.e., the binding is less tight. It seems the molecular polarity does not match well the surface electrostatic pattern. 
We note that the calculated binding energies are comparable to the ones previously reported for similar organic molecules (polythiophenes) on ZnO.\cite{doi:10.1021/jp201343s, doi:10.1021/jp203758p}

\begin{table} [ht]
\caption{Comparison of the binding free energy $\Delta F_{b} (z_{0})= \Delta U_{b} (z_{0})-T\Delta S_{b} (z_{0})$, energy of binding $\Delta U_{b} (z_{0})$ and the binding entropy contribution $T\Delta S_{b} (z_{0})$ between the {\it p}-6P,
{\it p}-6P2F and {\it p}-6P4F at T=525 K, where $z_{0}$ is equal to 0.184 nm, 0.195 nm and 0.202 nm for {\it p}-6P, {\it p}-6P2F and {\it p}-6P4F, respectively.}
\begin{tabular}{cccc}
\hline 
Molecule & {\it p}-6P & {\it p}-6P2F & {\it p}-6P4F\tabularnewline
\hline 
$\Delta F_{b} (z_{0}) $(kJ/mol)  & -97.48 & -93.75 & -87.02\tabularnewline

$\Delta U_{b} (z_{0}) $ (kJ/mol) & -164.97 & -132.61 & -122.01\tabularnewline

$T\Delta S_{b} (z_{0}) $ (kJ/mol) & -67.68 & -40.75 & -35.98\tabularnewline
%p-6P (\cite{doi:10.1021/jp507776h}) & 137.00 & 107.00 & 30.00\tabularnewline
\label{Table_bind}
\end{tabular}
\end{table}

\subsection{Translational diffusion on the surface}
\begin{figure}[ht]
\includegraphics[width=8.0cm]{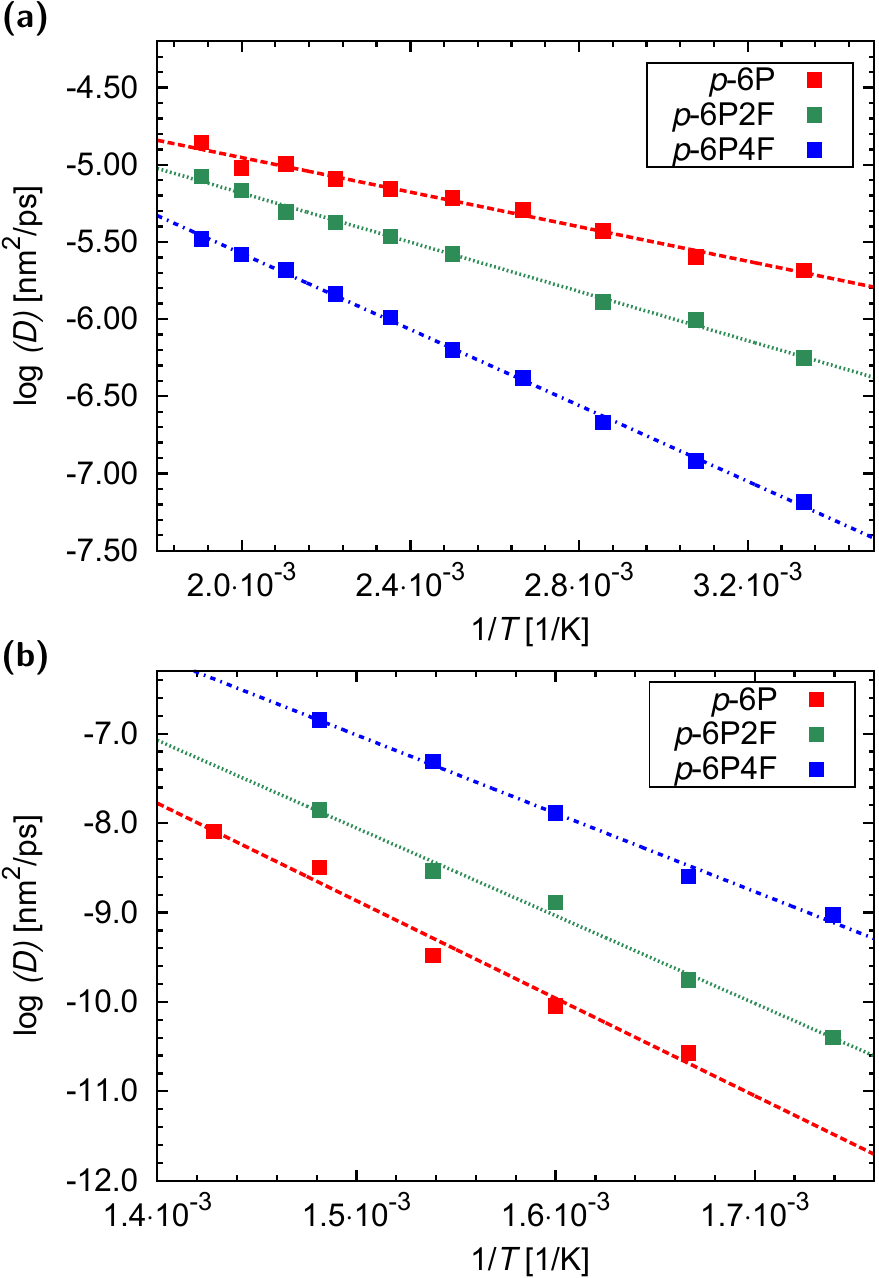}
\caption{
a) Logarithm of the diffusion coefficients in $x$ as a function of the inverse temperature. 
b) Logarithm of the diffusion coefficients in $y$ as a function of the inverse temperature. 
From the linear fits of the temperature dependent diffusion coefficients to the Arrhenius eq. (\ref{eq2}), the effective energy barriers $\Delta E_{\alpha}$ in $x$- and $y$-directions are deduced.}
\label{Ed_xy}
\end{figure}

We present now the results of single molecule diffusion in the non-polar $x$-direction. The time evolution of the molecular motion was simulated for 800 ns.
The temperature-dependent diffusion coefficients in $x$ (see Table S1 of the Supporting Information) are calculated according to eq.~(\ref{MSD}) from the mean squared displacements of the molecular COM (see Fig. S2 of the Supporting Information).
The corresponding diffusion coefficients are plotted in Fig. \ref{Ed_xy} a) as functions of inverse temperature. 
We find that they obey the Arrhenius law:
\begin{equation}
D(T) = D_{0}\exp\left(-\frac{\Delta E_{x}}{k_{B}T}\right),
\label{eq2}
\end{equation}
where $D_{0}$ is a prefactor and $\Delta E_{x}$ is the activation energy for the diffusion process in $x$-direction. 
By fitting the diffusion coefficients to the Arrhenius equation, we obtain the energy barrier $\Delta E_{x}$ for free diffusion from the slopes of the fits.
The energy barrier increases with increasing degree of fluorination from 4.65 to 6.95 to 10.67 kJ/mol for {\it p}-6P, {\it p}-6P2F and {\it p}-6P4F, respectively. 
The calculated $\Delta E_{x}$ of 4.65 kJ/mol for the {\it p}-6P molecule is about 4 times lower than the previously calculated value of 20 kJ/mol.~\cite{doi:10.1021/jp507776h}
However, in Ref.~\cite{doi:10.1021/jp507776h} it is also demonstrated that, when employing the full PME summation procedure as we do in this work, the barrier decreases to the value of about 6 kJ/mol, comparable to the $\Delta E_{x}$ calculated here. 
This exemplifies the sensitivity of the diffusion to small changes in molecule-surface interactions. 
We note that the average timescale for diffusion processes to be activated in $x$-direction to 'hop' one atomic step on the surface is about 10 ps for all investigated molecules (see Fig. S2 of the Supporting Information).

We separately evaluate the motion in the polar $y$-direction of the surface (see Table S2 and Fig. S3 of the Supporting Information for the $y$-components of the mean squared displacements of the molecular COMs).
We find that the average timescale for diffusion processes to be activated in $y$-direction on the surface is much larger than in $x$ and is of the order of 1 ns on average. 
From the slopes of the Arrhenius fits we deduce the energy barriers for free diffusion, $\Delta E_{y}$, in $y$-direction.  Opposite as in $x$-direction, they decrease with an increase in the degree of fluorination, with values of 90.45, 80.62, 69.20 kJ/mol for {\it p}-6P, {\it p}-6P2F and {\it p}-6P4F, respectively. 

Hence, we find the diffusion process to be strongly anisotropic for all investigated molecules, in which the diffusive motion along the polar [0001] direction of the surface is many orders of magnitude slower than in the perpendicular direction. 
%All molecules showed anisotropic diffusion behavior with a preference to diffuse along the $x$-direction most of the time, while the diffusion in y was less preferred, requiring higher temperatures and longer simulation times for the random, uncorrelated jumps in y to result in a long-time diffusion process. 
Interestingly, the molecule with the highest number of fluorine atoms has the lowest barrier for $y$-diffusion but the highest barrier for $x$-diffusion.   This happens because of the electrostatic repulsion between the negatively charged oxygen atoms of the surface and also negatively charged fluorine atoms of the molecule. At the same time, zinc and fluorine atoms attract each other due to the different charge sign. The net effect of the interplaying repulsion and attraction causes the barrier in $y$ to be lower for the {\it p}-6P4F molecule.

We note again that the total friction in our model that acts on the molecule during its diffusion on the surface has two independent contributions (see Methods \ref{sec:Diffusion}), namely the intrinsic molecule-surface contribution and the auxiliary bath friction. 
The intrinsic one $\xi_\alpha^{ms}$ is the one of interest and simply follows as $\xi_\alpha^{ms}=\xi_\alpha^{\rm tot}-\xi_\alpha^{mb}$, where $\alpha$=$x$, $y$. The results we just discussed are qualitatively the same as for the total diffusion (see Fig. S4  b) and c) and Fig. S5 a)-d) of the Supporting Information). 

%Fig. S4 b) and c) also indicate that {\it p}-6P and {\it p}-6P2F experience about 2 orders of magnitude higher friction when moving in the polar $y$-direction, while for {\it p}-6P4F this difference is of about one order of magnitude between the two directions.
% Same as D, nothing new here..

\subsection{Free energy landscapes for the diffusion on the surface}

\begin{figure}[ht]
\includegraphics[width=7cm]{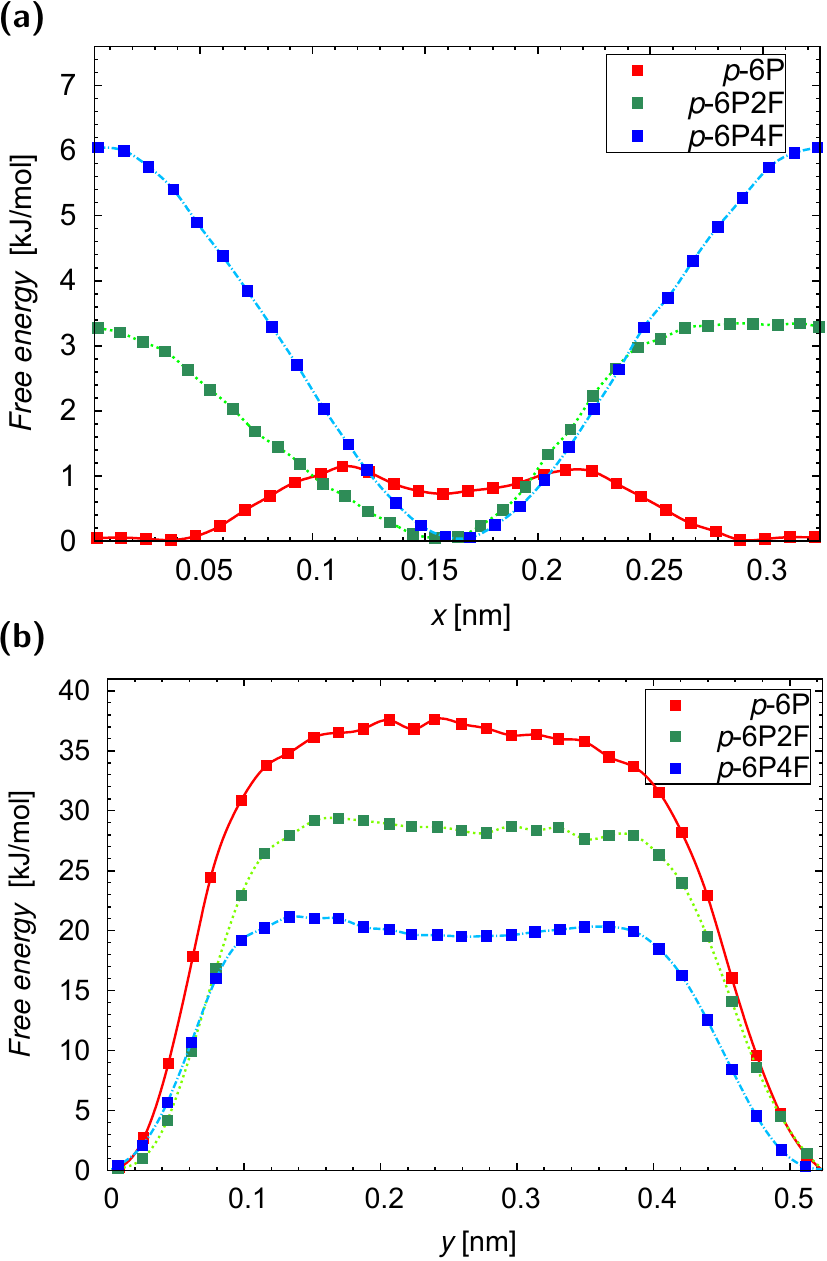}
\caption{Free energy as a function of a) $x$- and b) $y$-position of the molecular COM at T=625 K, calculated using the eq.~(\ref{eq3}).
The free energy barrier, $\Delta F_{\alpha}$, is defined as the difference between the maximum and the minimum point of the curve (see Methods \ref{sec:Free_en}).}
\label{Fig_freexy}
\end{figure}

%In order to calculate the free energy landscapes of the molecules, as well as to quantify entropic contributions to the free energies, we extract from the simulation trajectories the probabilities of presence $P(x)$ and $P(y)$ for the molecular COM simply from Boltzmann inversion (see Methods).  
The detailed diffusion free energy landscapes will tell us more about the conformity or mismatch of molecular and surface polarity. They are calculated according to the description in the Methods section \ref{sec:Free_en}. In Fig.~\ref{Fig_freexy} we see that the three molecules have very different free energy profiles for the diffusion along $x$. 
The peak value in the free energy curve in $y$ for the {\it p}-6P (35.5 kJ/mol) is slightly higher than the peak value in Ref.~\cite{doi:10.1021/jp507776h} (31 kJ/mol).
Also, the peak value in $x$ (1.1~kJ/mol) is about 2~kJ/mol lower than the previously calculated value in Ref.~\cite{doi:10.1021/jp507776h} (3.2 kJ/mol). 
We again attribute this difference to the different treatment of the long-range electrostatic interactions in the system, which has an effect on the diffusion energy barriers, as already discussed above. The free energy barriers increase with fluorination in $x$, while they decrease in $y$. 
In Table~\ref{Table4} we separate the entropic and energetic parts of the diffusion free energy barriers and compare the calculated values between the three investigated molecules. The potential energy barriers are in good agreement with the diffusion energy barriers derived from the Arrhenius analysis in Fig. 4, demonstrating consistency of the independent approaches and the correctness of the results. 

We can see that the diffusion free energy barriers have significant entropic components in both the $x$- and $y$-direction. The total entropic barrier, $T \Delta S$, may be separated into contributions that appear due to the translation and rotation on the surface but also contributions that come from the internal molecular degrees of freedom.
In case of the {\it p}-6P, we find the entropy contribution coming from the internal molecular motion (averaged over all COM positions in $x$ and $y$) to be higher than in the cases of {\it p}-6P2F and {\it p}-6P4F (see Figure S6 of the Supporting Information). 
The increase in configurational entropy associated with inter-ring torsion helps the {\it p}-6P to reduce the free energy barriers for surface diffusion, similarly as observed for the crossing of an atomic surface step-edge barrier.~\cite{C6CP05251G, Zhang2018}

\begin{table} [ht]
\caption{Energetic $\Delta E_{\alpha}$ and entropic contribution $T \Delta S_{\alpha}$ to the free energy barrier $\Delta F_{\alpha}$ between the {\it p}-6P, {\it p}-6P2F and {\it p}-6P4F resolved in $\alpha=x,y$-direction of the motion.}
\centering
\begin{tabular}{cccc}
%\hline
molecule & $\Delta E_{x}$ (kJ/mol) & $T \Delta S_{x}$ (kJ/mol) & $\Delta F_{x}$ (kJ/mol)\tabularnewline
\hline 
%\hline 
{\it p}-6P & 5.24 & 4.11 & 1.13\tabularnewline

{\it p}-6P2F & 8.80 & 5.48 & 3.32\tabularnewline

{\it p}-6P4F & 14.04 & 8.00 & 6.04\tabularnewline
\hline
\hline
%\hline 
%\hline 
\end{tabular}
\begin{tabular}{cccc}
 & $\Delta E_{y}$ (kJ/mol) & $T\Delta S_{y}$  (kJ/mol) & $\Delta F_{y}$  (kJ/mol)\tabularnewline
\hline 
{\it p}-6P & 89.05 & 53.55 & 35.50\tabularnewline

{\it p}-6P2F & 79.51 & 51.39 & 27.72\tabularnewline
 
{\it p}-6P4F & 66.49 & 44.97 & 21.52\tabularnewline
\hline 
\hline 
\label{Table4}
\end{tabular}
\end{table}

\begin{figure*}[ht!]
\includegraphics[width=13cm]{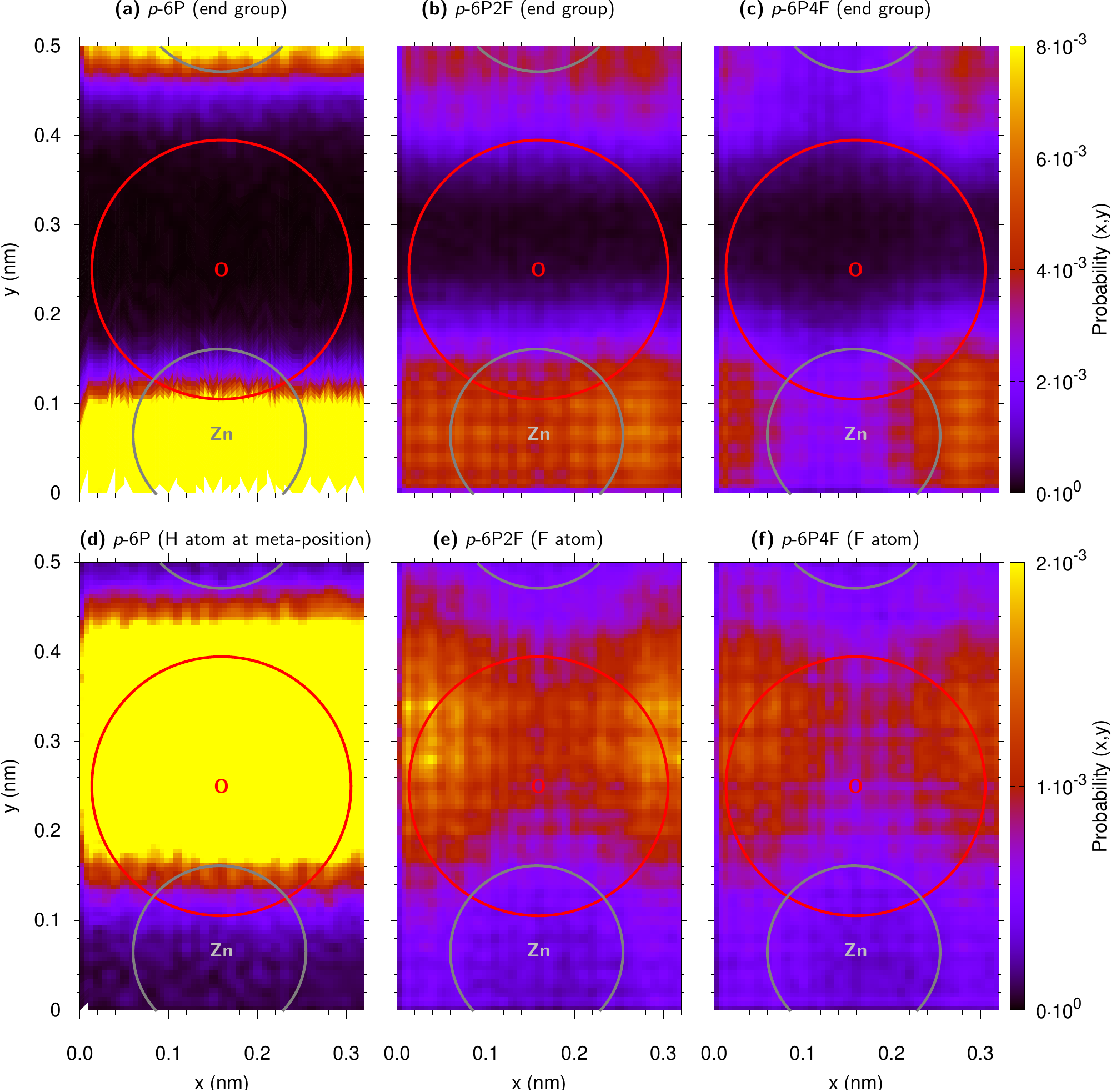}
\caption{Top panels: Probability (see color-bar) as a function of $x$- and $y$-positions for the COM of the end-phenyl-group in the a) {\it p}-6P, b) {\it p}-6P2F and c) {\it p}-6P4F molecule. 
Bottom panels: Probability (see color-bar) as a function of $x$- and $y$-positions of the end-group H atom in the a) {\it p}-6P and end-group F-atoms of the b) {\it p}-6P2F and c) {\it p}-6P4F molecules. 
The radii of the plotted circles represent the vdW radii of zinc and oxygen atoms, taken from Ref. \cite{JCC:JCC20035}.}
\label{P_FH_endgroup}
\end{figure*}  

In Fig. \ref{P_FH_endgroup} a), b) and c) we plot the two-dimensional probability distributions $P(x,y)$ for the end-group COM position of each molecule on the surface.
The data are folded onto one ZnO unit-cell.
The fluorinated end-groups of {\it p}-6P2F and {\it p}-6P4F demonstrate a higher rotational and translational freedom and a more evenly distributed probability to sample positions in the $y$-direction compared to the end-groups of the {\it p}-6P molecule.
Fig.~\ref{P_FH_endgroup} d), e) and f) shows the same kind of probability distribution but for the position of a meta-hydrogen atom in the {\it p}-6P case and a fluorine atom in the case of {\it p}-6P2F and {\it p}-6P4F. 
Again, the probability for the fluorinated molecules is much more evenly distributed than for {\it p}-6P.
Due to the net effect of attractive F-Zn and repulsive F-O interactions, the fluorine atoms prefer the positions between the oxygen atoms of the surface during the simulation while the meta-hydrogen atoms of the {\it p}-6P end-groups strongly prefer to stay above the oxygen atoms.
Even though the end-groups of the {\it p}-6P2F and {\it p}-6P4F are chemically and structurally the same, they sample slightly different positions on the surface unit-cell, as visible in Fig. \ref{P_FH_endgroup} b) and c) as well as e) and f).
This is attributed to the different number of fluorinated groups in each molecule: Since {\it p}-6P2F has both, a fluorinated and a normal phenyl head group it sample states that are a mix of Fig. \ref{P_FH_endgroup} a) and c) as well as d) and f).

\subsection{Angular motion and rotational diffusion on the surface}
\begin{figure*}[ht!]
\includegraphics[width=13cm]{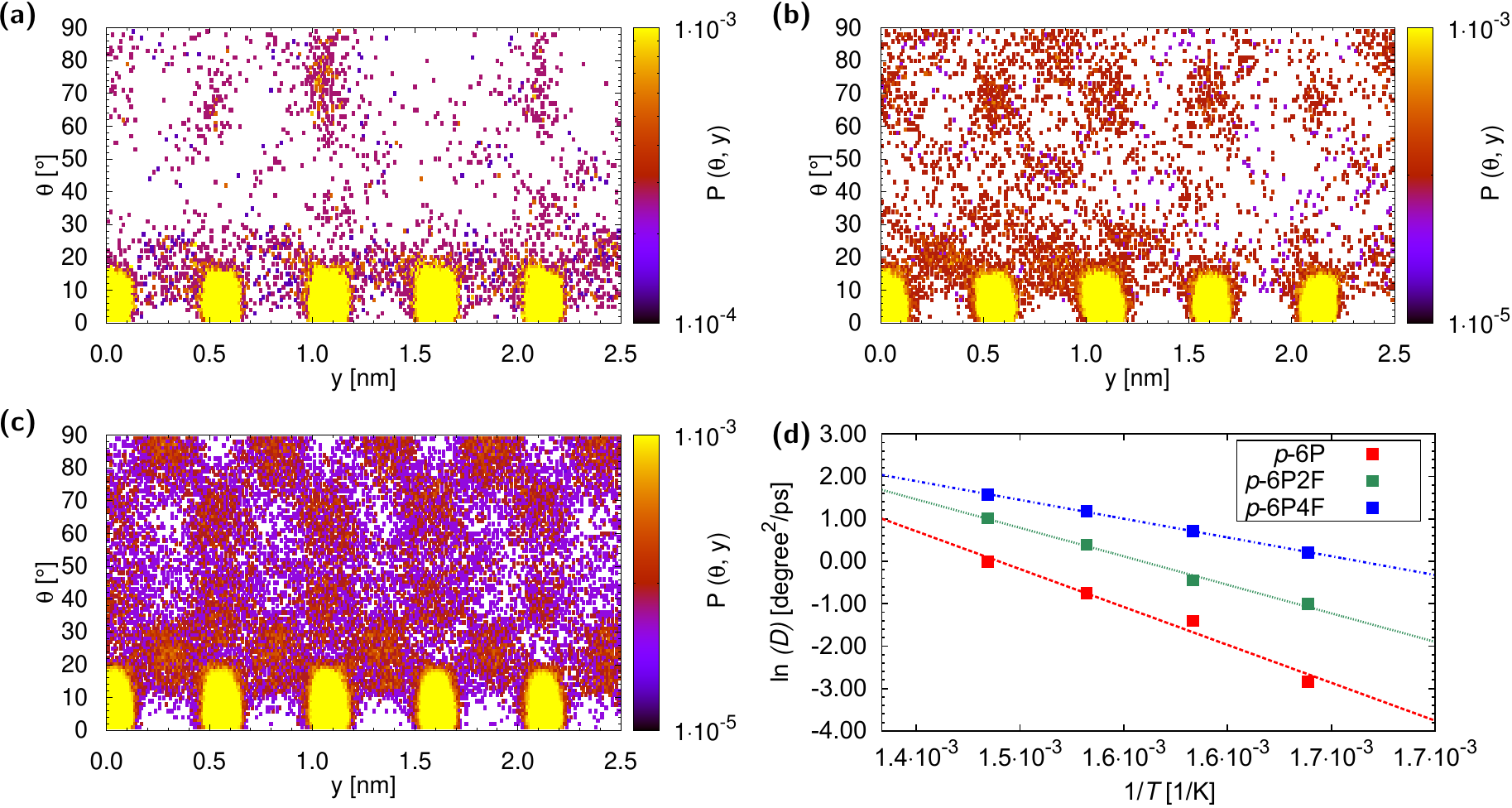}
\caption{Probability densities (see color-bar) as a function of the orientational angle ${\theta}$ and position in $y$ for a) {\it p}-6P, b) {\it p}-6P2F and {\it p}-6P4F at $T = 600$~K. d) Arrhenius plots of rotational diffusion coefficients as a function of inverse temperature for the three investigated molecules.}
\label{Rot_ed2}
\end{figure*} 
We finally study the rotational diffusion of the molecules by calculating the in-plane orientational angle, ${\theta}$, of the LMA to the $x$-direction of the surface in the temperature range of 600 K to 675 K. 
%Since the molecules prefer to align their LMAs parallel with the $x$-direction most of the time, temperature plays an essential role for the molecular in-plane rotation: in the studied thermal range the surface-molecule binding free energy is smaller than at lower temperatures. 
The relatively high temperatures are necessary for the molecules to sample the orientational conformations during the 800~ns long simulations. 
Fig.~\ref{Rot_ed2} depicts the stable (preferred) orientational and positional coordinates for the molecules on the surface at $T = 600$~K. 
While the {\it p}-6P4F has the lowest activation barrier for the movement in $y$, Fig.~\ref{Rot_ed2} shows that it has the lowest activation barrier for free rotation, too.
Comparing panels a), b) and c), we can see that {\it p}-6P4F is able to find more energetically favorable pathways in the $y$ direction compared to the other molecules. 
The stable states on the surface are separated by well defined periodic distances and the most sampled conformations are the ones where the molecules are aligned with their LMAs almost parallel to the $x$-axis.
Configurations on the surface where the LMAs are oriented with ${\theta} = 90$\textdegree~to the $x$-axis are energetically unfavorable and rarely sampled in case of {\it p}-6P and also {\it p}-6P2F. 

The angular motion of the investigated molecules is thermally activated and characterized by corresponding diffusion coefficients (see Table \ref{D_rot}), that are calculated from linear fits of the time dependent mean squared angular displacements (see Fig. S7 a), b) and c) of the Supporting Information).
From the temperature dependence of the rotational diffusion coefficients, we deduce the rotational activation energies and list them in Table~\ref{Table6}. 
From Fig.~\ref{Rot_ed2} a), b) and c) we can see that all molecules rotate approximately $20 \pm 5$\textdegree~before they ascend the barrier and move to the next stable position on the surface. 
By rotating the molecule away from a low energy configuration on the surface, the (free) energy barrier for translation decreases, making it easier for the molecule to translate in $y$.  
The coupling between the different degrees of freedom of adsorbed molecules is a phenomenon already observed and characterized for other molecular adsorbates.~\cite{C6CP05251G, doi:10.1063/1.465508} 
\begin{table}
\caption{Rotational diffusion coefficients (in \textdegree/ps) between the {\it p}-6P,
{\it p}-6P2F and {\it p}-6P4F.}
\begin{tabular}{cccc}
$T$ (K) & {\it p}-6P &  {\it p}-6P2F &  {\it p}-6P4F\tabularnewline
\hline 

\hline 
600     & 0.058 & 0.363 & 1.232    \tabularnewline
625     & 0.244 &  0.638 & 2.037 \tabularnewline
650     & 0.472 & 1.475 & 3.252\tabularnewline
675     & 0.986 &  2.757 & 4.796 \tabularnewline
\hline 
\hline 
\label{D_rot}
\end{tabular}
\end{table}

\begin{table}
\caption{Rotational energy barriers of the {\it p}-6P,
{\it p}-6P2F and {\it p}-6P4F.}
\begin{tabular}{cccc}
molecule & {\it p}-6P &  {\it p}-6P2F &  {\it p}-6P4F\tabularnewline
\hline 
\hline 
$\Delta E_{a}$  (kJ/mol) & 123.71 & 93.05 & 61.27\tabularnewline
\hline 
\hline 
\label{Table6}
\end{tabular}
\end{table}

\section{Summary and Concluding Remarks}
In summary, we have theoretically investigated the role of polarity of single {\it p}-6P, {\it p}-6P2F and {\it p}-6P4F molecules in the binding and diffusion on a ZnO (10\=10) surface.  DFT and MD calculations of the molecular total dipole moment in the gas-phase revealed a permanent dipole in the {\it p}-6P2F molecule of the order of 3 Debye, in both planar and twisted geometries. In  the anti-symmetric {\it p}-6P4F two local dipoles of such magnitude cancel each other.  It is then observed that the heterogeneous electrostatic surface pattern of the (10\=10) ZnO surface causes a highly anisotropic diffusion with very different energy barriers for the three investigated molecules. 
We have shown that the increase in the number of fluorinated, polar groups decreases the diffusivity in the non-polar $x$-direction of the surface, but increases the diffusivity in the polar $y$-direction. This unexpected behavior of different trends with fluorination in different directions could be traced back to complex electrostatic many-body interactions between the negatively charged fluorine atoms on the one hand, and the positively charged zinc atoms and negatively charged oxygen atoms on the other hand. As a net effect there is an electrostatic mismatch that leads to overall weaker binding of the more fluorinated molecules and facilitates the diffusion in $y$ and the rotation on the surface plane. This behavior was explained in detail by detecting the driving molecule-surface interactions that govern the diffusion process in the polar $y$-direction. 

An important implication of our findings is that partial fluorination of the {\it p}-6P molecule can significantly alter its surface binding and surface diffusion on a single molecule level by modifying the degree of anisotropy in the net effective surface free energy landscape. Clearly the specific electrostatic pattern of the surface plays a decisive role. This has implications for the rational design of molecules and their functionalized forms which could be tailored for a programmable anisotropic match or mismatch between molecular polarity and electrostatic surface patterns.  

The complex interplay between the molecule-molecule repulsive and attractive interactions has already been demonstrated in experimental epitaxy to have a significant effect in molecular self-assembly on surfaces.~\cite{doi:10.1021/acs.jpcc.8b03398} 
Our study fully supports this view and provides unprecedented in-depth details of the molecular realization of this interplay between the molecule-surface attractive and repulsive interactions. The detailed understanding is crucial for  the control of molecular self-assembly behavior and, thus, needs to be further examined experimentally to enable an improved design of molecular self-assembly, nucleation and growth.  

Regarding theoretical multi-scale modeling approaches to self-assembly and growth,  our work provides the parameters that are required in large scale kinetic Monte Carlo simulations, such as diffusion coefficients and energy barriers.~\cite{PhysRevE.98.042801}
With proper molecule-molecule and surface-molecule interactions as input, self-assembly and growth can the be simulated for experimentally relevant scales.~\cite{doi:10.1021/acs.cgd.6b00109, C4CP04048A}

\section{Acknowledgements}
Funded by the Deutsche Forschungsgemeinschaft (DFG, German Research Foundation) - Projektnummer 182087777 - SFB 951 (project A7). 
The authors wish to thank Sabine Klapp and Thomas Martynec for inspiring discussions.

\newpage

\bibliography{acs-achemso}
\clearpage
\newpage

\begin{figure}
\includegraphics[width=17.8cm]{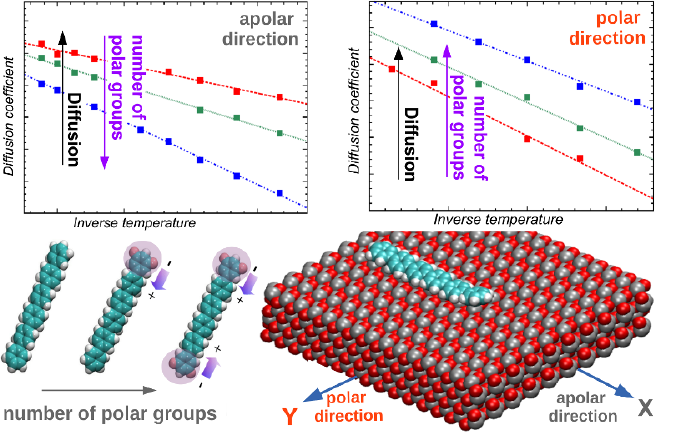}
\caption{Table of Contents Figure.}
\end{figure}

\end{document}